\documentclass[useAMS,usenatbib]{aa}
\usepackage{txfonts}
\usepackage{natbib}
\bibpunct{(}{)}{;}{a}{}{,}
\usepackage{longtable}
\usepackage[dvips]{graphicx}
\begin{document}
\title{Three X-ray transients in M~31 observed with Swift}
\author{R. Voss\inst{1} \and W. Pietsch\inst{1} \and F. Haberl\inst{1} 
\and H. Stiele\inst{1} \and J. Greiner\inst{1} \and G. Sala\inst{1} \and D.H. Hartmann\inst{2} \and D. Hatzidimitriou\inst{3}}
\institute{Max-Planck-Institut f\"ur extraterrestrische Physik, 
Giessenbachstrasse, 85748 Garching, Germany
\and Department of Physics and Astronomy, Clemson University, Clemson, SC 29634
\and University of Crete, Department of Physics, PO Box 2208, 71003 Heraklion, Greece
\\
email:rvoss@mpe.mpg.de}
\titlerunning{X-ray transients in M~31}
\date{Received ... / Accepted ...}

\offprints{R. Voss}

\abstract{}
{
The purpose of this study is to find transient X-ray sources in M~31, and
to investigate and classify their nature.
}
{
Three X-ray transients were observed with \textit{Swift}.
For each of the three X-ray transients we use the \textit{Swift}
X-ray and optical data together with observations from \textit{XMM-Newton}
and \textit{Chandra} to investigate the lightcurves and the spectra of the
outburst, and thereby to identify the source types.
}
{
The outburst of XMMU J004215.8+411924 lasted for about one month. 
The source had a hard power-law spectrum with a photon index of 1.6.
It was previously identified as a Be/X-ray binary based on
the optical identification with a star. However, we show
that with improved source coordinates it is clear that the optical
source is not the counterpart to the X-ray source.
The source SWIFT J004217.3+411532 had a bright outburst, after which it
slowly decayed over half a year. The spectrum was soft, corresponding
to a thermal accretion disk with innermost temperature of 
$\sim250-600$ eV. The source was
not seen in the optical, and the soft spectrum indicates that the source 
is most likely a black hole low mass X-ray binary. M31N 2006-11a 
is a nova that was previously observed in the
optical. We detected it both in X-rays and UV with 
\textit{Swift} $\sim$half a year after the optical maximum, after
which it decayed below the \textit{Swift} detection threshold within
a month. The spectrum of the X-ray transient can be modelled by a black-body
with a temperature of 50 eV. We use catalogues of X-ray transients in
M~31 to estimate their rate, and we find a lower limit of 9 yr$^{-1}$.
}
{}


\maketitle

\section{Introduction}
Being the nearest large spiral galaxy, M~31 is an important target for the
study of X-ray point sources in galaxies. In a large number of studies,
X-ray point sources in M~31 have been investigated and catalogued, using
\textit{Einstein} \citep{Trinchieri}, \textit{ROSAT} \citep{Primini,Supper}, 
\textit{XMM-Newton} \citep[e.g.][]{Trudolyubov2,Pietsch} and 
\textit{Chandra} \citep[e.g.][]{Kong,Kaaret,Williams1,Williams2,Voss}.\\
An interesting subset of these sources are the transients. These are sources
with quiescent luminosities below the detection limit 
($\sim 10^{35}$ erg s$^{-1}$) that display bright outbursts typically 
lasting from weeks to a few months. From Galactic observations it is
well-known that there are three types of objects that can show such 
outbursts. Optical studies of soft X-ray transients have shown that
they contain an accreting compact object more massive
than 3 $M_{\odot}$, revealing the presence of
black hole accretors \citep{McClintock}. In contrast the harder sources
typically show pulsations, indicating that the accreting object is
a neutron star \citep[e.g.][]{Tanaka,Campana}. The third class is
the thermal emission from classical novae in optical decline, showing
up as super-soft X-ray transients \citep{MacDonald,Starrfield,Pietsch2007}.  
Monitoring programs have shown that 
6--12 transients appear per year in the bulge of M~31 \citep{Trudolyubov}, 
and within the timescale
of the X-ray observations, only a small fraction of them have been 
observed to recur.\\
Here we report on the detection and analysis of three transient 
X-ray sources in the bulge of M~31, 
using data from \textit{Swift},
\textit{Chandra} and \textit{XMM-Newton}. 

\begin{table*}
\begin{center}
\caption{The list of the observations used for the analysis
of the three X-ray transients, TR1 (XMMU J004215.8+411924), 
TR2 (SWIFT J004217.4+411532) and TR3 (M31N 2006-11a).
In the columns of the three sources, a
"+" signifies that the source was detected in this observation, whereas
a "-" signifies that it was not detected.}
\label{tab:obs}
\begin{tabular}{llccccccc}
\hline\hline
Observation &Date & Observatory & Instrument & Obs-ID & TR1 & TR2 & TR3\\
\hline
1 &2006 Jun 05 & Chandra & HRC-I & 7283       & - & - & -\\
2 &2006 Jul 02 & XMM     & EPIC  & 0405320501 & - & - & -\\
3 &2006 Jul 31 & Chandra & ACIS-I& 7139       & + & - & -\\	
4 &2006 Aug 09 & XMM     & EPIC  & 0405320601 & + & - & -\\
5 &2006 Sep 01 & Swift   & XRT   & 00030802001& + & + & -\\
6 &2006 Sep 11 & Swift   & XRT   & 00030804001& - & + & -\\
7 &2006 Sep 24 & Chandra & ACIS-I& 7140     &-&+& \\
8 &2006 Sep 30 & Chandra & HRC-I & 7284	   & - & + & -\\
9 &2006 Nov 13 & Chandra & HRC-I & 7285       & - & + & -\\
10& 2006 Dec 4 & Chandra & ACIS-I & 7064 &  & + & \\
11&2006 Dec 31 & XMM     & EPIC  & 0405320701 & - & + & -\\
12&2007 Jan 16 & XMM     & EPIC  & 0405320801 & - & + & -\\
13&2007 Mar 11 & Chandra & HRC-I & 7286       & - & - & -\\
14&2007 Jun 01 & Swift   & XRT   & 00035336001& - & - & +\\
15&2007 Jul 12 & Swift   & XRT   & 00030968001& - & - & -\\
16&2007 Jul 13 & Swift   & XRT   & 00030968002& - & - & -\\
17&2007 Jul 18 & Swift   & XRT   & 00030968004& - & - & -\\
\hline
\end{tabular}
\end{center}
\end{table*}
\section{Data analysis}
\label{sect:data}
A 20 ks \textit{XMM-Newton} observation on 2006, August 9 as part of
the monitoring program of the X-ray supersoft state of the optical novae
in the core of 
M~31
revealed a new bright transient, designated 
XMMU J004215.8+411924 \citep{Haberl}\footnote{{\tiny http://www.mpe.mpg.de/xray/research/normal\_galaxies/index.php}}.
In a 6 ks follow-up target of opportunity observation with \textit{Swift}
on 2006, September 1, the source was seen again \citep{Pietsch_Atel}. 
A second bright transient was detected in this observation and was
designated SWIFT J004217.4+411532. A third source was found in
a 19.4 ks \textit{Swift} observation 2007, 
June 1 \citep{Pietsch_Atel2}. We identify this source
as optical nova M31N 2006-11a, the optical outburst of which was
detected about 190 days earlier\footnote{{\tiny http://cfa-www.harvard.edu/iau/CBAT\_M31.html\#2006-11a}}.\\
In Table \ref{tab:obs} we list the observations analysed in this
study. The distance to M~31 was
assumed to be 780~kpc, and all fluxes and
luminosities are given in the 0.5--5~keV range, except for the
nova M31N 2006-11a for which the luminosities are given in the
0.1--1.0 keV range. All statistical
errors and upper limits are given with 90\% confidence.
The data were reduced using 
{\tt SAS} version 7.1.0
for \textit{XMM-Newton} observations, 
{\tt CIAO} version 3.4 
for the \textit{Chandra} observations and 
{\tt HEAsoft} version
6.3 for the \textit{Swift} observations.
The spectra were modelled with {\tt Xspec}, using the {\tt tbabs}
absorption model with the abundance tables of \citet{Wilms} and the
photoelectric absorption cross-sections of \citet{Balucinska} and \citet{Yan}.
For \textit{XMM-Newton} we performed simultaneous fits to the
spectra from the pn and the two MOS detectors, except for observations
where the source was in or near to detector gaps.
All luminosities
are given as the intrinsic source luminosities, corrected
for absorption.
\section{The Sources}
\subsection{XMMU J004215.8+411924}
\label{sect:XMM}
\begin{figure}
\resizebox{\hsize}{60mm}{\includegraphics[angle=0]{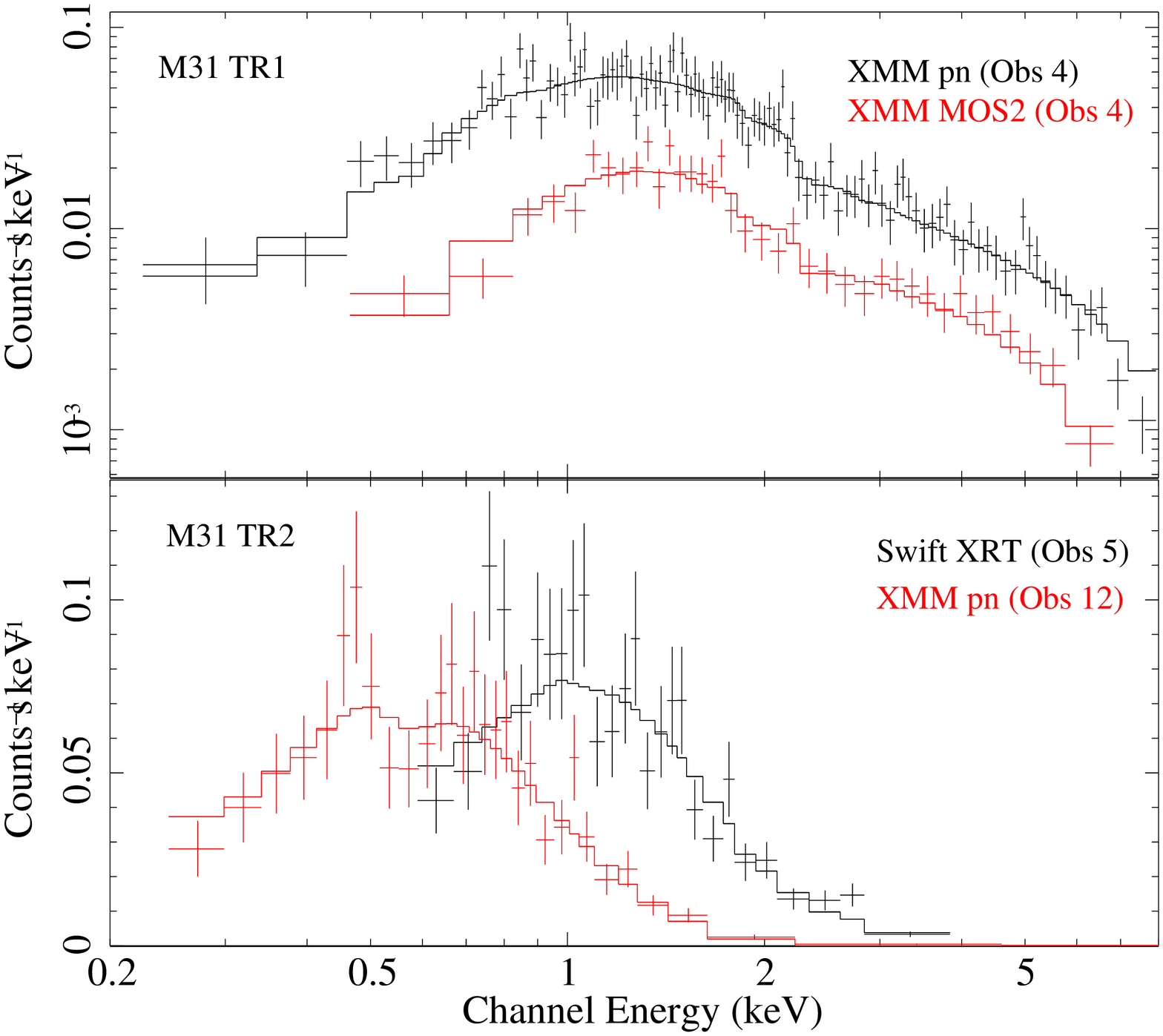}}
\caption{The spectrum of XMMU J004215.8+411924 (top) from observation 4, 
with absorbed power-law fits. The spectrum
obtained from the pn (black) and the MOS2 (red) detectors are shown.
The source is located in a gap on the MOS1 detector. 
The spectrum of SWIFT J004217.4+411532 (bottom) from observations 5 (highest
observed luminosity, black) and 12 (lowest observed luminosity, red), 
with absorbed disk black-body fits.}
\label{fig:TR1}
\end{figure}

\begin{table*}
\caption{The parameters resulting from the fits to the spectra
of XMMU J004215.8+411924 and SWIFT J004217.4+411532. The luminosities are given as 
$L_{36}=L/10^{36}$~erg~s$^{-1}$, and the neutral hydrogen column
is given by $N_{H,21}$=$N_{H}/10^{21}$~cm$^{-2}$, $\Gamma$ is the photon
index of the power-law fits, and the temperatures $T_{in}$ and $T_b$
of the disk black-body and the bremsstrahlung fits are given in keV.
Parameters in parentheses were kept fixed at the assumed values and
were not the result of fits to the data.}
\label{tab:TR1}
{\tiny
\begin{center}
\begin{tabular}{cccccccccccccc}
\hline
\hline
& \multicolumn{4}{c}{power-law}& \multicolumn{4}{c}{Disk black-body} &\multicolumn{4}{c}{Bremsstrahlung} \\
\hline
Obs & $L_{36}$ & $\Gamma$ & $N_{H,21}$ & $\chi^2_\nu$ & $L_{36}$ & $T_{in}$ & $N_{H,21}$ & $\chi^2_\nu$
& $L_{36}$ & $T_b$ & $N_{H,21}$ & $\chi^2_\nu$\\
\hline
\multicolumn{14}{c}{XMMU J004215.8+411924}\\
\hline
2 & $<0.4$ & (1.57) & (4.2) & -& $<3.5$& (1.92) & (2.2) & -& $<3.8$& (12.2) & (3.6)& -\\
3 & $110\pm30$ & 1.77$\pm0.33$ & 6.2$\pm$3.7 & 0.98 &
$86\pm44$ & 1.55$\pm$0.35 & 2.5$\pm$2.3 & 1.05&
$100\pm50$ & 6.5$\pm$4.0 & 5.2$\pm$2.7 & 0.98&
\\
4 & $76\pm12$ & 1.57$\pm0.08$ & 4.2$\pm$0.5 & 0.84 &
$67\pm11$ & 1.92$\pm$0.14 & 2.2$\pm$0.3 & 0.87&
$73\pm12$ & 12.2$\pm$4.0 & 3.6$\pm$0.4 & 0.80&\\
5 & 21$\pm$1 & (1.57) & (4.2) & -& 19$\pm1$& (1.92) & (2.2) & -&  20$\pm$1& (12.2)& (3.6)&-\\
6 & $<5.8$ & (1.57) & (4.2) & -& $<5.1$& (1.92) & (2.2) & -&  $<5.6$&(12.2) & (3.6) & -\\
7 & $<0.8$ & (1.57) & (4.2) & - & $<0.7$ & (1.92) & (2.2) & - & $<0.6$ & (12.2) & (3.6) & -\\
\hline
\multicolumn{14}{c}{SWIFT J004217.4+411532}\\
\hline
4 & $<2.1$ & (3.5) & (4.0) & - & 
$<0.9$ & (0.50) & (0.67) & - & $<0.9$ & (1.0) & (1.5) & -\\
5 & $580\pm70$ & $3.12\pm0.35$ & $4.4\pm1.3$ & 1.35 & 
$270\pm30$ & $0.61\pm0.05$ & (0.67) & 1.26 & 350$\pm70$ & 1.26$\pm0.26$ & 2.18$\pm0.91$ & 1.28\\
6 & $450\pm70$ & $3.55\pm0.49$ & $4.0\pm1.7$ & 0.99
& $180\pm30$ & $0.50\pm0.06$ & (0.67) & 0.96 &230$\pm70$& 1.07$\pm0.27$& 1.13$\pm1.13$ & 0.86 \\
7  & $260\pm60$ & $3.47\pm0.31$ & $5.1\pm1.3$ & 0.87 &  $ 97\pm7$ & 0.57$\pm0.04$ & (0.67) & 0.97&  $130\pm20$ &$1.15\pm0.11$  & $2.0\pm0.9$ & 0.88\\
8 & $290\pm10$ & (3.5) & (4.0) & - & $ 140\pm10$ & (0.50) & (0.67) & - & 122$\pm4$ & (1.0) & (1.5) & -\\
9 & $163\pm7$ & (3.5) & (4.0) & - & $ 80\pm4$ & (0.50) & (0.67) & - & $69\pm3$ & (1.0) & (1.5) & -\\
10& $81\pm29$ & $4.50\pm0.65$ & $5.1\pm2.2$ & 1.26 &  $20\pm2$ & 0.35$\pm0.03$ & (0.67) & 1.01& $28\pm15$ & 0.58$\pm0.12$ & 1.91$\pm1.33$ & 1.10\\
11 & $36\pm4$ & $4.34\pm0.33$ & $3.3\pm0.5$ & 1.56 &
$15\pm3$ & $0.27\pm0.02$ & (0.67) & 1.67 & 18$\pm4$ & 0.49$\pm$0.06 & 1.49$\pm0.27$ & 1.52\\
12 & $20\pm3$ & $4.51\pm0.57$ & $3.2\pm0.8$ & 1.10&
$8.2\pm2.2$ & $0.25\pm0.03$ & (0.67) & 1.44 & 10$\pm$4 & 0.42$\pm$0.08 & 1.50$\pm0.43$ & 1.29\\
13 & $<0.9$ & (3.5) & (4.0) & - &
$<0.5$ & (0.50) & (0.67) & - & $<0.4$ & (1.0) & (1.5) & -\\
\hline
\end{tabular}
\end{center}
}
\end{table*}
As is shown in Table \ref{tab:obs} this source was active in
three X-ray observations (Observations 3 to 5) spanning 40 days, and
from the non-detections in observation 2 and 6, the maximum 
duration of the outburst can be constrained to 79 days.
In Table \ref{tab:TR1} the results
of modelling the source spectra with three different models,
absorbed power-law, disk black-body and bremsstrahlung are
shown. The data can be equally well represented by each of
the models, and the obtained luminosities agree within the
errors. In the following analysis we use the power-law
interpretation, due to the relatively hard spectrum
(power-law photon index $\Gamma\simeq1.6$), noting that none
of our conclusions are affected by the choice of model.
The spectrum obtained from observation 4 is shown in Fig. \ref{fig:TR1},
together with the best fitting power-law model.\\

\begin{figure}
\resizebox{\hsize}{55mm}{\includegraphics[angle=0]{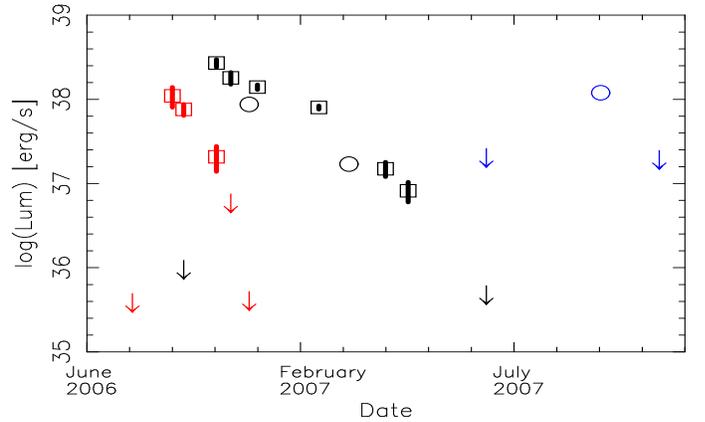}}
\caption{The light curves of the three sources, XMMU J004215.8+411924 (red),
SWIFT J004217.4+411532 (black) and M31N 2006-11a (blue). The arrows
signify upper limits, the squares indicate data points where
fits with spectral information were employed, and the circles correspond
to luminosities estimated from spectra with model parameters fixed at 
expected values.}
\label{fig:curves}
\end{figure}

In observation 5 the number of source counts is too low to fit
a model to the data, and we used the spectrum from observation 4
to estimate the luminosity of the
source, $2.1(\pm0.7)\cdot10^{37}$~erg~s$^{-1}$, a factor of $\sim$4 lower
than in the previous observations.
We used observations 2 and 6 to calculate upper limits on the
quiescent luminosity of the source, assuming the spectral shape found
in observation 4.
These upper limits are given in Table \ref{tab:TR1}
and the lightcurve of the transient is shown in Fig. \ref{fig:curves}.
It can be seen that the source is variable by
a factor of more than $\sim300$, with a peak luminosity of
$\sim10^{38}$~erg~s$^{-1}$. Galactic binaries often show variability
factors of $10^5-10^6$, but with the sensitivity limits in M~31,
only lower limits on the variability factor of 100-1000 can be found.\\

\begin{figure}
\resizebox{\hsize}{!}{\includegraphics[angle=0]{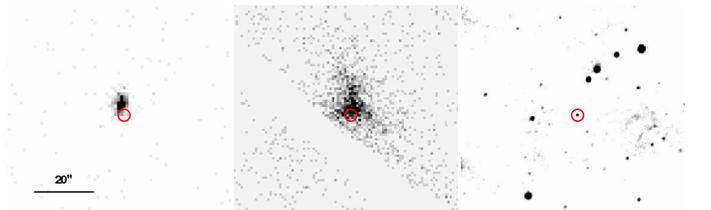}}
\caption{The position of XMMU J004215.8+411924 in X-rays and optical.
Observation 3 with \textit{Chandra} ACIS-I (left), observation
4 with MOS2 (middle), and the \textit{B}-band image from the
LGS (right). The red circle has a radius of 2 arcsec and marks the 
previously identified optical counterpart
to the X-ray source.}
\label{fig:TR1_2}
\end{figure}

The source was previously identified as a Be/X-ray binary, based on
the identification of an optical counterpart observed with the 
\textit{Swift} UVOT \citep{Haberl}. With the available \textit{Chandra}
data the position of the source can be improved significantly.
We derive the source position from observation (3), giving 
R.A.(J2000)=00:42:16.1, Dec.(J2000)=+41:19:26.7, 
with a 1$\sigma$ error of 0.5 arcsec, including the statistical
error on the position, as well as the error on the boresight
of the observation, estimated from cross-correlation of the source
positions with 2MASS \citep{Skrutskie} sources.\\
We reinvestigated the correlation between the optical and the X-ray source,
and we furthermore investigated the images from the Local Group Survey
(LGS) of \citet{Massey}. In Fig. \ref{fig:TR1_2} we show the region of the
source in X-rays and optical.
There is an offset between
the optical source and the position of the \textit{Chandra} X-ray source of 
$\sim4$~arcsec. This is much larger than the position error, and
it shows that the match between the X-ray source and the
optical counterpart was coincidental.
We therefore conclude that the proposed optical source is not the counterpart
to the X-ray source. However, the limiting magnitude of the optical 
observations is not high enough to rule out a Be companion star. For
example the limiting magnitude for this area of the LGS in the
\textit{V}-band is $\sim22$, and with a distance modulus of 24.46 and
an extinction of $\sim0.4$ \citep{Han}, Be stars can have magnitudes
as faint as 25 \citep{Wegner,Garmany}. On the other hand, the Be/X-ray
binaries mostly contain more luminous Be stars, and for example all
known Be/X-ray binaries in the Small Magellanic Cloud \citep{Coe}
are luminous enough that they would be observed in the optical if they
were at the position XMMU J004215.8+411924.
The spectrum and lightcurve of the outburst is consistent with a
Be/X-ray binary in a type~II outburst. However, in the absence of
a detectable optical counterpart it is not possible to confirm this.
\subsection{SWIFT J004217.4+411532}
This transient was first observed in the \textit{SWIFT} follow-up
observation of XMMU J004215.8+411924 on 2006, September 1.
It stayed in outburst until after January 2007, and it
was observed 8 times (observations 5 to 12). The inferred duration
of the outburst is between 125 and 175 days.
For the observations 5, 6, 11 and 12 we extracted source spectra 
and performed fits using the same three models as for 
XMMU J004215.8+411924.
The results of this analysis are shown in Table \ref{tab:TR1}.
The spectrum is very soft and the three models fit the
spectra equally well. For X-ray binaries in outburst,
such a soft spectrum is always found to be emission from an
accretion disk, and we therefore use the disk black-body
interpretation in the following analysis.
For this model, all fits yielded
neutral hydrogen columns consistent with just the Galactic foreground
absorption, and the fits were repeated with the column density fixed to
this value. 
Due to the lack of spectral information from 
\textit{Chandra} HRC, we used fixed models (with the parameters given
in parenthesis in Table \ref{tab:TR1}) to convert the count
rates to fluxes in observations 8 and 9. Finally we used the
same fixed models to place upper limits on the luminosity of
the source in observations 4 and 13. The
light-curve is displayed in Fig. \ref{fig:curves}.\\

After a fast rise to the
maximum luminosity of $\sim3\cdot10^{38}$~erg~s$^{-1}$, the
source luminosity decayed exponentially to a quiescent level
a factor of $>500$ lower than the peak luminosity, with an
e-folding timescale $\tau_e\sim30-40$ days. According to the
absorbed disk black-body fits
the inner disk temperature decreased from 0.6~keV at the peak
luminosity to $\sim0.25$~keV. In Fig. \ref{fig:TR1} a comparison
between the spectrum at the outburst peak and
at the lowest detected luminosity is shown.
The behaviour of the source is
typical of black hole candidate outbursts to the high/soft
state \citep{McClintock}.\\ 
We use observation 8 to
find the most precise position of the source, R.A.(J2000)=00:42:17.3,
Dec.(J2000)=+41:15:37.2 (1$\sigma$ error 0.5 arcsec).
We found no optical source at the X-ray position from searching 
the \textit{Swift} UVOT observations, as well as from \textit{Galex} and
\textit{HST} and the LGS observations. However, as for XMMU J004215.8+411924 the
observations are not sensitive enough to constrain the source type.\\
\subsection{M31N 2006-11a}
This optical nova was detected by K. Itagaki on
November 24th, 2006 and it reached 17.3 mag at maximum on unfiltered
CCD images\footnote{{\tiny http://cfa-www.harvard.edu/iau/CBAT\_M31.html\#2006-11a}}. 
The position was reported to be R.A.(J2000)=00:42:56.81,
Dec.(J2000)=+41:06:18.4.
From a spectral analysis of the optical nova, A. Shafter concluded that it
was of the Fe II type\footnote{{\tiny http://mintaka.sdsu.edu/faculty/shafter/extragalactic\_novae/HET/index.html}}. 
We observed a transient source in only one observation (14),
with \textit{Swift} at the position of the nova.
In this observation a soft X-ray source with about 26 counts is detected,
just inside the field of view of the XRT, with all
photons below 530 eV. The temperature of a black-body fit to the source
spectrum can be constrained to 40 eV$<$kT$<$ 90 eV, using the Galactic
foreground absorption as a lower bound and constraining the luminosity
to below $3.5\cdot10^{38}$~erg~s$^{-1}$ (the Eddington luminosity of
a white dwarf at the Chandrasekhar mass limit, assuming a He-rich
atmosphere).\\
Assuming a temperature of 50 eV and the best-fit absorbing column of
$1.1\cdot10^{21}$~cm$^{-2}$, we derive upper limits to the luminosity 
before and after the detection.
From observation 13 two months before the detection 
we derive an upper limit of 
$2.0\cdot10^{37}$~erg~s$^{-1}$.
From combining observations 15, 16 and 17 we place an upper limit of
1.9$\cdot10^{37}$~erg~s$^{-1}$ a month after the detection. The
resulting lightcurve is shown in Fig. \ref{fig:curves}.\\
During observation 14, a \textit{Swift} UVOT exposure of the nova region
of 4.2 ks in \textit{U} and 7.7 ks in UVW2 was obtained. The nova was clearly
detected in both filters at magnitudes 20.15$\pm$0.19 and 21.12$\pm$0.16,
respectively.
The emission in UVW2 (180$\pm$20 nm) may well result from the 
[CIII] 190.9 nm, [NIII] 175.0 nm and/or [OIII] 166.3 nm lines,
typical for the spectra of novae in the nebular phase.\\
We searched for the optical source in observations prior to the
detection. We did not detect the source in the \textit{Galex} image, 
nor in the existing \textit{HST} observations, and
the source was not inside the field of view of the UVOT in the
\textit{Swift} observations 5 and 6.\\ 
About 1.5 months after the X-ray detection, the source was detected in
the UVW2 band in observations 15 and 17 with magnitudes of 21.95$\pm0.30$
and 21.62$\pm0.19$, respectively and in the \textit{U}-band in
observation 16 with magnitude 20.63$\pm$0.11, corresponding to a
luminosity decrease of $\sim$0.5 mag
between observations 14 and 15. The source was not detected in
later observations.\\
While the X-ray luminosity dropped by a factor of 10 between
observations 14 and 17, the \textit{UV} luminosity only changes
by a factor of 2.
This indicates that the strong decrease in the flux of observed 
X-rays is due to a decrease in temperature of the emitting region.
Assuming a temperature of
50 eV for the source in observation 14, and an absorbing column of 
1.1$\cdot10^{21}$ cm$^{-1}$, the upper limit on the X-ray to 
\textit{UV} ratio in observation 15 to 17 corresponds to a black-body
temperature below 37 eV.

\section{Discussion}
To compare the sources discussed in this paper with previously observed
sources and to understand how the fit into recognized classes
of sources, we select a sample from the two catalogues of transient X-ray
sources in M~31 \citep{Williams2,Voss}, based on the criteria that a
variability of $>20$ is established, and that spectral information must
be available.
This sample consists of 43 transients, observed over
a period of 4.6 yr (October 1999 -- May 2004) in 45 \textit{Chandra}
observations as well as with 5
epochs of \textit{XMM-Newton} data. As the spatial and temporal
coverage is far from complete from these observations, we can set
a lower limit to the rate of transients of $\sim 9$ transients yr$^{-1}$,
revising the lower limit of 6 transients yr$^{-1}$ obtained by \citep{Trudolyubov}
somewhat upwards.\\
\begin{figure}
\resizebox{\hsize}{!}{\includegraphics[angle=0]{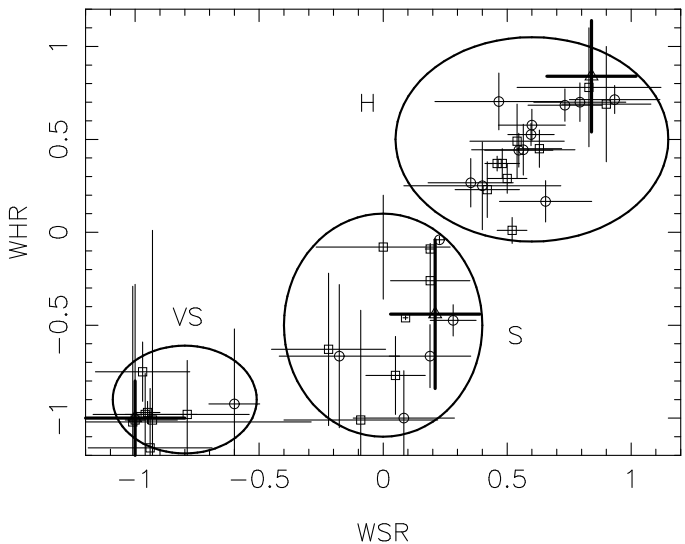}}
\caption{The hardness ratios at the highest observed luminosities 
of 41 transients observed in M~31. The sources are divided into three
subcategories: Very soft (VS), soft (S) and hard (H). Open squares
are data points from \citep{Williams2}, while circles are data points
from our analysis of transients listed in \citet{Voss}. The three
sources analysed in this paper are marked with triangles and have
thick error-bars. The displayed
errors correspond to 1-sigma confidence intervals.}
\label{fig:colours}
\end{figure}
We classified the sources, using the hardness ratios of the sources
as given in \citet{Williams2}\footnote{WSR=(WM-WS)/(WM+WS) and WHR=(WH-WS)/(WH+WS), where WS, WM and WH are
the source counts in the 0.3--1.0, 1.0--2.0 and 2.0--7.0 keV band,
respectively. Note that these ratios are
different from the hardness ratios defined in section \ref{sect:XMM}}.
We characterized the outburst by the hardness ratios at
the highest observed luminosity. For the transients published in 
\citet{Williams2} we used their values, whereas
for the sources in \citet{Voss},
we analysed the \textit{Chandra} observations to find the hardness
ratios. As shown in Fig. \ref{fig:colours} the sources can be naturally divided
into three distinct populations (with the three sources discussed in this paper
fitting into one class each): the very soft transients, the soft
transients, and the hard transients. From this analysis we found that half
(20) of the observed sources belong to the hard class, with the other
half distributed evenly between the soft (12) and the very soft (9)
sources.

\section{Conclusions}
For the first time, \textit{Swift} was used to observe X-ray transients in
M~31. We analysed three sources, two of which were discovered from the \textit{Swift}
observations, thereby showing the possibilities of using \textit{Swift} for such
studies. We furthermore made use of all available \textit{Chandra} and 
\textit{XMM-Newton} observations to analyse the developments in luminosity and
spectrum for each of the sources.
We investigated the rate of transients in M~31 and we found a
minimum rate of 9 yr$^{-1}$, and that $\sim$half of the sources
have a hard spectrum.

\begin{acknowledgements}
This research has made use of data obtained from the Chandra Data Archive 
and software provided by the Chandra X-ray Center (CXC) in the application 
package CIAO and observations obtained with XMM-Newton, 
an ESA science mission with instruments and contributions directly funded by ESA Member States and NASA.
The XMM-Newton project is supported by the Bundesministerium f\"ur 
Wirtschaft und 
Technologie/Deutsches Zentrum f\"ur Luft- und Raumfahrt (BMWI/DLR, FKZ 
50 OX 0001), 
the Max-Planck Society. We thank the 
\textit{Swift}
team for their help with scheduling the TOO observations.
We have made use of observations made with the NASA/ESA Hubble Space 
Telescope and with the Galaxy Evolution Explorer, obtained from the 
data archive at the Space Telescope Institute. 
STScI is operated by the association of Universities for Research in 
Astronomy, Inc. under the NASA contract  NAS 5-26555.
G.S. and H.S. are supported through DLR (FKZ 50 OR 0405).
\end{acknowledgements}


\begin{thebibliography}{}
\bibitem[\protect\citeauthoryear{Arnaud}{1996}]{Arnaud}
Arnaud, K.A., 1996, Astronomical Data Analysis Software and Systems V, 
eds. Jacoby G. and Barnes J., ASP Conf. Series volume 101, p17
\bibitem[\protect\citeauthoryear{Balucinska-Church \& McCammon}{1992}]{Balucinska} 
Balucinska-Church, M., \& McCammon, D., 1992, ApJ, 400, 699
\bibitem[\protect\citeauthoryear{Campana et al.}{1998}]{Campana} 
Campana, S., Colpi, M., Mereghetti, S., Stella, L., Tavani, M., 1998,
A\&ARv, 8, 279
\bibitem[\protect\citeauthoryear{Coe et al.}{2005}]{Coe} 
Coe, M.J., Edge, W.R.T., Galache, J.L., McBride, V.A. 2005, MNRAS, 356
\bibitem[\protect\citeauthoryear{Garmany \& Humphreys}{1985}]{Garmany}
Garmany, C.D., \& Humphreys, R.M., 1985, AJ, 90, 2009
\bibitem[\protect\citeauthoryear{Haberl et al.}{2006}]{Haberl}
Haberl, F., Pietsch, W., Greiner, J., Ajello, M., \& the collaboration
to monitor the supersoft state of optical novae in M~31, 2006, ATel, 881
\bibitem[\protect\citeauthoryear{Han}{1996}]{Han} Han, C. 1996,
ApJ, 472, 108
\bibitem[\protect\citeauthoryear{Kaaret}{2002}]{Kaaret} Kaaret, P. 2002,
ApJ, 578, 114
\bibitem[\protect\citeauthoryear{Kong et al.}{2003}]{Kong} Kong, A.K.H.,
Di Stefano, R., Garcia, M.R., Greiner, J. 2003, ApJ, 585, 298
\bibitem[\protect\citeauthoryear{MacDonald \& Vennes}{1991}]{MacDonald}
MacDonald, J., \& Fujimoto, M.Y., Truran, J.W. 1985, ApJ, 294, 263
\bibitem[\protect\citeauthoryear{McClintock \& Remillard}{2006}]{McClintock}
McClintock, J.E., Remillard, R.A. 2006, in Compact stellar X-ray sources,
ed. W. Lewin \& M. van der Klis, Cambridge University Press, Cambridge, UK
\bibitem[\protect\citeauthoryear{Massey et al.}{2006}]{Massey}
Massey, P., Olsen, K.A.G., Hodge, P.W., Strong, S.B., Jacoby, G.H., 
Schlingman, W., Smith, R.C. 2006, AJ, 131, 2478 
\bibitem[\protect\citeauthoryear{Pietsch et al.}{2005}]{Pietsch}
Pietsch, W., Fliri, J., Freyberg, M.J., Greiner, J., Haberl, F., Riffeser, A.,
Sala, G. 2005a, A\&A, 442, 879
\bibitem[\protect\citeauthoryear{Pietsch et al.}{2006}]{Pietsch_Atel}
Pietsch, W., Haberl, F., Greiner, J., Stiele, H. 2006, ATel 899
\bibitem[\protect\citeauthoryear{Pietsch et al.}{2007a}]{Pietsch_Atel2}
Pietsch, W., Greiner, J., Haberl, F., Sala, G. 2007a, Atel, 1116
\bibitem[\protect\citeauthoryear{Pietsch et al.}{2007b}]{Pietsch2007}
Pietsch, W., Haberl, F., Sala, G., et al. 2007b, A\&A 465, 375
\bibitem[\protect\citeauthoryear{Primini, Forman \& Jones}{1993}]{Primini}
Primini, F.A., Forman, W., \& Jones, C. 1993, ApJ, 410, 615
\bibitem[\protect\citeauthoryear{Skrutskie et al.}{2006}]{Skrutskie}
Skrutskie, M.F., Cutri, R.M., Stiening, R., et al. 2006, AJ, 131, 1163
\bibitem[\protect\citeauthoryear{Starrfield}{1989}]{Starrfield}
Starrfield, S. 1989, in "Classical Novae", ed. M. Bode \& A. Evans
(Wiley, New York), 39
\bibitem[\protect\citeauthoryear{Supper et al.}{2001}]{Supper}
Supper, R., Hasinger, G., Lewin, W.H.G., Magnier, E.A., van Paradijs, J.,
Pietsch, W., Read, A.M., Tr\"umper, J., 2001, A\&A, 373, 63
\bibitem[\protect\citeauthoryear{Tanaka \& Shibazaki}{1996}]{Tanaka}
Tanaka, Y., \& Shibazaki, N. 1996, ARA\&A, 34, 607
\bibitem[\protect\citeauthoryear{Trinchieri \& Fabbiano}{1991}]{Trinchieri}
Trinchieri, G., \& Fabbiano, G. 1991, ApJ, 382, 82
\bibitem[\protect\citeauthoryear{Trudolyubov et al.}{2002}]{Trudolyubov2}
Trudolyubov, S.P., Borozdin, K.N., Priedhorsky, W.C., Mason, K.O., 
Cordova, F.A., 2002, ApJ, 571, 17
\bibitem[\protect\citeauthoryear{Trudolyubov et al.}{2006}]{Trudolyubov}
Trudolyubov, S., Priedhorsky, W., Cordova, F., 2006, ApJ, 645, 227 
\bibitem[\protect\citeauthoryear{Voss \& Gilfanov}{2007}]{Voss}
Voss, R., \& Gilfanov, M. 2007, A\&A, 468, 49
\bibitem[\protect\citeauthoryear{Wegner}{2000}]{Wegner}
Wegner, W., 2000, MNRAS, 349, 193
\bibitem[\protect\citeauthoryear{Williams et al.}{2004}]{Williams1} Williams,
  B.F., Garcia, M.R., Kong, A.K.H., Primini, F.A., King, A.R., Di Stefano, R.,
Murray, S.S. 2004, ApJ, 609, 735
\bibitem[\protect\citeauthoryear{Williams et al.}{2006}]{Williams2}
Williams, B.F., Naik, S., Garcia, M.R., Callanan, P.J. 2006, ApJ, 643, 356
\bibitem[\protect\citeauthoryear{Wilms et al.}{2000}]{Wilms}
Wilms, J, Allen, A., McCray, R., 2000, ApJ, 542, 914
\bibitem[\protect\citeauthoryear{Yan et al.}{1998}]{Yan}
Yan, M., Sadeghpour, H. R., Dalgarno, A., 1998, ApJ, 496, 1044
\end{thebibliography}
\end{document}